\documentclass[11pt]{iopart} 
\usepackage{epsfig,bm} 
\def \cD{{\cal D}} 
\def \cN{{\cal N}} 
\def \cO{{\cal O}}

\def\bbbone{{\rm 1\hspace{-1.1mm}I}}	

\def \rd {\mbox{\rm d}}

\def \vu {{\bm u}}

\def \be{\begin{equation}} 
\def \ee{\end{equation}}  
\def \bea{\begin{eqnarray}}
\def \eea{\end{eqnarray}}

\begin{document} 
\title{Spectra of Modular Random Graphs} 
\author{G\"uler Erg\"un$^1$ and Reimer K\"uhn$^2$}
\address{$^1$Department of Mathematical Sciences, University of Bath,\\
Claverton Down, Bath BA2 7AY, UK\\[1mm]
$^2$Mathematics Department, King's College London, Strand, London WC2R 2LS,UK} 
\begin{abstract}
We compute spectra of symmetric random matrices defined on graphs exhibiting a 
modular structure. Modules are initially introduced as fully connected sub-units
of a graph. By contrast, inter-module connectivity is taken to be incomplete.
Two different types of inter-module connectivity are considered, one where the
number of intermodule connections per-node diverges, and one where this number 
remains finite in the infinite module-size limit. In the first case, results 
can be understood as a perturbation of a superposition of semicircular spectral
densities one would obtain for uncoupled modules. In the second case, matters 
can be more involved, and depend in detail on inter-module connectivities. For 
suitable parameters we even find near-triangular shaped spectral densities, 
similar to those observed in certain scale-free networks, in a system of 
consisting of just two coupled modules. Analytic results are presented for
the infinite module-size limit; they are well corroborated by numerical
simulations.
\end{abstract}
\section{Introduction}
\label{sec:intro}

Appreciation has steadily grown in recent years that theories of networked
systems provide useful paradigms to understand complex processes 
in various branches of science and technology, including the evolution 
of the internet and the world-wide web, flows of information, power, or
traffic, credit, market, or operational risk, food-webs in ecosystems, 
gene-regulation, protein-protein interactions underlying metabolic 
processes or cell signalling, immune system response, information 
processing in neural networks, opinion formation, the adoption of new 
technologies in societies, or the spread of diseases or epidemics, and
more (see, e.g. \cite{AlbBarab02, DorogMend03} for recent reviews).

Random matrix theory has long been known to constitute a powerful tool
to study topological properties of the graphs underlying networked
systems \cite{Cvetk+95, Bollobas01, Farkas+01, AlbBarab02, Dorog+03}.
E.g., moments of the spectral density of an adjacency matrix describing 
a graph, gives complete information about the number of walks returning
to the originating vertex after a given number of steps. 

Any form of heterogeneity existing in a graph, be it due to a scale-free
distribution of connectivities \cite{BarabAlb99}, due to a small-world
structure \cite{WaStro98}, or due to the existence of a modular structure 
in terms of identifiable sub-graphs \cite{MitTad08}, or a certain density 
of specific motifs existing in a graph would, therefore, manifest itself
in the spectral density of states of the corresponding connectivity matrix
\cite{Cvetk+95,  Bollobas01, Farkas+01, AlbBarab02, Dorog+03, MitTad08}. 
A triangular density of states, for instance, has been hailed as a signature 
of a scale-free connectivity distribution emerging from network growth by 
preferential attachment \cite{Farkas+01}. 

Spectra of complex networks with heterogeneity arising from scale-free
or small-world connectivity patterns were obtained through numerical 
diagonalization \cite{Farkas+01, Goh+01}; analytical results for certain 
networks with power-law degree distributions exist, based either on
the effective medium approximation(EMA) \cite{NagRog08}, or restricted to
the limit of large average connectivity \cite{Rodg+05,KimKahng07}. A complete 
solution for general degree distributions with finite average connectivity
(scale-free or other) has been obtained only recently, both for ensembles 
in the thermodynamic limit \cite{Ku08}, and for large single instances 
\cite{Rog+08}, the latter generalised to non-hermitean matrices in 
\cite{RogPer+09}. A highly efficient approximation for Poissonian random 
graphs at moderately large mean connectivity has appeared in \cite{Dea02}. 
We are not aware of studies other than numerical for the case of networks 
with small world \cite{Farkas+01} or explicitly modular structure 
\cite{MitTad08}.

In the present paper, we introduce a block-structured random matrix model 
to study spectral properties of coupled systems, or of a system with modules.
The model can be thought of as a network composed of sub-networks, with
weighted links within and between subnetworks. To simplify matters, we will
initially consider the case where modules are defined as fully connected
subnetworks. However, we note at the outset that this case is not 
substantially different from the one where intra-module connectivity is 
incomplete but with coordination that diverges in the infinite module-size 
limit. Inter-module connectivity is assumed to be incomplete, and two 
substantially different cases will be considered, one where the number 
of inter-module connections per node diverges, and one where this number 
remains finite in the infinite module-size limit. 

We shall find that results for the first case can be rationalized in terms
of perturbed superpositions of semi-circular spectral densities characteristic
of the individual modules when considered on their own, whereas in the second
case matters can be more involved, and depend in greater detail on inter-module
connectivities. For suitable parameters we even find near-triangular shaped
spectral densities, similar to those observed in certain scale-free networks,
already for a case of two coupled modules.

The remainder of the paper is organized as follows. In Sec. 2 we present an
elementary variant of a modular system a  system consisting of two coupled
modules. This section mainly serves to introduce the formal definitions, and
the method we will be using for computing the density of states, viz. the 
the replica method as proposed in the random matrix context by Edwards and 
Jones \cite{EdwJon76}. Details of the replica calculation for the extensively
cross-connected case are given in Sec. 2.1. The fundamentally different
case of (two) finitely cross-connected modules is dealt with in Sec. 2.2. In
Sec. 3 we present a generalisation of the self-consistency equations derived
earlier for the general $M$-modules case. Results for the extensively and finitely
cross-connected two-modules case are presented and discussed in Secs 4.1 and
4.2, respectively. The paper concludes with a summary and outlook in Sec. 5.

\section{A Modular System}
We begin by considering the most elementary modular system described by a $2N\times 2N$
symmetric block matrix of the form 
\be
H = \left(\begin{array}{cc} M^{(1)} &V\\V^t& M^{(2)}\end{array}\right)
\ee
in which the symmetric $N\times N$ sub-matrices $M^{(1)}$, $M^{(2)}$, and the 
$N\times N$ coupling-sub-matrix $V$ are random. We assume that elements of
$M^{(1)}$  and $M^{(2)}$ are independently and identically Gaussian distributed:
$M^{(\mu)}_{ij} \sim \cN(0,J_\mu^2/N)$, $1\le i\le j\le N$,
 $\mu=1,2$, and that
$V_{ij}=c_{ij} K_{ij}$ with $K_{ij} \sim \cN(0,J_p^2/c)$, $1\le
 i, j\le N$, while
\be
p(c_{ij})=\left(1-\frac{c}{N}\right) \delta_{c_{ij},0} + \frac{c}{N} \delta_{c_{ij},1}\ .
\ee
Two different limits will be considered: (i) the so-called extensively cross-connected
limit $c\to\infty$, $N\to\infty$, with the ratio $c/N$ either remaining finite or
approaching zero in the $N\to\infty$-limit.\footnote{The second alternative could more
appropriately be referred to as sub-extensive, a distinction we are not going to make
here for simplicity, as it does not affect the nature of results} (ii) the so-called
finitely cross-connected limit, where $c$ is kept constant, as $N\to\infty$. In this case
the distribution of inter-module connectivities is Poissonian in the thermodynamic limit,
with average coordination $c$.

We are interested in the spectral density of $H$,
\be
\rho_N(\lambda) = \frac{1}{2N} \sum_{k=1}^{2N} \delta(\lambda-\lambda_k)\ ,
\ee
more precisely in its average $\overline{\rho_N(\lambda)}$ over the random matrix ensemble 
introduced above. Here, the $\lambda_k$ are the eigenvalues of $H$. We shall use 
$\overline{\rho(\lambda)}$ to denote the (average) spectral density in the thermodaynamic
limit,
\be
\overline{\rho(\lambda)}= \lim_{N\to\infty} \overline{\rho_N(\lambda)}\ .
\ee

The spectral density is computed from the resolvent via
\be
\rho_N(\lambda) = \lim_{\varepsilon\searrow 0} ~ \frac{-2}{2N\pi}{\rm Im} 
\frac{\partial}{\partial \lambda} \ln {\rm det}\left[\lambda_\varepsilon\bbbone
-H \right]^{-1/2}\ ,
\ee
in which $\lambda_\varepsilon\equiv\lambda- i\varepsilon$, and the inverse square root of
the determinant is obtained as a Gaussian integral.  Using $\vu^{(1)}$ and $\vu^{(2)}$ to
denote $N$ component vectors, and $\vu=(\vu^{(1)},\vu^{(2)})$ to denote their
concatenation, we get 
\be
\overline{\rho_N(\lambda)}= \lim_{\varepsilon\searrow 0} ~ \frac{-2}{2N\pi}{\rm Im} 
\frac{\partial}{\partial \lambda} \left \langle \ln \left[
\int \frac{\rd\vu^{(1)} \rd \vu^{(2)}}{(2\pi/i)^N} \exp\left\{- \frac{i}{2} \vu \cdot 
\left[\lambda_\varepsilon \bbbone -H\right]\vu\right\}\right] \right\rangle\ ,
\ee
where angled brackets on the r.h.s denote an average over connectivities $\{c_{ij}\}$ and
weights $\{M^{(\mu)}_{ij}\}$ and $\{K_{ij}\}$ of the non-vanishing matrix elements.

The average of the logarithm is evaluated using replica.
\be
\overline{\rho_N(\lambda)}= \lim_{\varepsilon\searrow 0} ~ \frac{-2}{2N\pi}{\rm Im} 
\frac{\partial}{\partial \lambda}\lim_{n\to 0}\frac{1}{n} \ln \langle Z_N^n \rangle\ ,
\label{AvDOSn}
\ee
with
\be
Z_N^n=\int \prod_a\frac{\rd\vu_a^{(1)} \rd \vu_a^{(2)}}{(2\pi/i)^N} \exp\left\{-
 \frac{i}{2} \sum_{a=1}^n \vu_a
\cdot \left[\lambda_\varepsilon \bbbone - H\right]\vu_a\right\}\ .
\ee
Here $a=1,\dots,n$ enumerates the replica.
\subsection{The Extensively Cross-Connected Case}

The average $\langle Z_N^n \rangle$ is easily performed \cite{EdwJon76, RodgBray88}. We
have
\bea
\langle Z_N^n \rangle \!\!\! &=& \!\!\!\int \prod_a\frac{\rd\vu_a^{(1)} 
\rd \vu_a^{(2)}}{(2\pi/i)^N} 
\exp\left\{-\frac{i}{2} \sum_a \lambda_\varepsilon (\vu^{(1)}_a \cdot \vu^{(1)}_a +
\vu^{(2)}_a \cdot \vu^{(2)}_a )\right\} \nonumber\\
& &\times
\left \langle \exp\left\{ \frac{i}{2} \sum_a\vu^{(1)}_a \cdot M^{(1)} 
\vu^{(1)}_a \right\}  \right\rangle \times
\left \langle \exp\left\{ \frac{i}{2} \sum_a\vu^{(2)}_a \cdot M^{(2)} \vu^{(2)}_a \right\}
\right\rangle\nonumber\\
& &\times
\left \langle \exp\left\{ i \sum_a\vu^{(1)}_a \cdot V\vu^{(2)}_a \right\}\right\rangle 
\nonumber\ .
\eea
Up to subdominant corrections from diagonal matrix elements this gives
\bea
\langle Z_N^n \rangle &\simeq& \int \prod_a\frac{\rd\vu_a^{(1)} \rd \vu_a^{(2)}}{(2\pi/i)^N}
\exp\left\{- \frac{i}{2} \sum_a \lambda_\varepsilon (\vu^{(1)}_a \cdot \vu^{(1)}_a +
\vu^{(2)}_a \cdot \vu^{(2)}_a )\right .\nonumber\\
&& \left . -\frac{J_1^2}{4N} \sum_{a,b} (\vu^{(1)}_a \cdot \vu^{(1)}_b)^2
	   -\frac{J_2^2}{4N} \sum_{a,b} (\vu^{(2)}_a \cdot \vu^{(2)}_b)^2
\right\}\nonumber\\
& & \times \prod_{ij}\left[1 + \frac{c}{N}\left(\exp\left\{-\frac{J_p^2}{2c}
\sum_{a,b} u^{(1)}_{ia}u^{(1)}_{ib}u^{(2)}_{ja}u^{(2)}_{jb}\right\}-1 \right)
\right]
\label{eq:avzn}
\eea
Expanding the exponential in the product (for large $c$) and re-exponentiating one
obtains
\bea
\hspace{-10mm}\langle Z_N^n \rangle &\simeq& \int \prod_a\frac{\rd\vu_a^{(1)} \rd \vu_a^{(2)}}{(2\pi/i)^N}
\exp\left\{- \frac{i}{2} \sum_a \lambda_\varepsilon (\vu^{(1)}_a \cdot \vu^{(1)}_a +
\vu^{(2)}_a \cdot \vu^{(2)}_a )-\frac{J_1^2}{4N} \sum_{a,b} (\vu^{(1)}_a \cdot \vu^{(1)}_b)^2 \right .\nonumber\\
& & \left . 
	   -\frac{J_2^2}{4N} \sum_{a,b} (\vu^{(2)}_a \cdot \vu^{(2)}_b)^2
	   -\frac{J_p^2}{2N} \sum_{a,b} (\vu^{(1)}_a \cdot \vu^{(1)}_b)(\vu^{(2)}_a \cdot
	   \vu^{(2)}_b)
\right\}
\eea
Decoupling of sites is achieved by introducing 
\be
q^{(\mu)}_{ab} =\frac{1}{N}
\sum_i u^{(\mu)}_{ia}u^{(\mu))}_{ib} ~~~~~ ,\quad \mu=1,2\ ,
\ee
as order parameters and by enforcing their definition via $\delta$-functions. This leads
to
\be
\langle Z_N^n \rangle \simeq \int \prod_{\mu,a,b}\frac{\rd q^{(\mu)}_{ab} \rd\hat 
q^{(\mu)}_{ab} }{2\pi/N}~\exp\{N[G_1 + G_2 + G_3]\}
\label{ZNSPInt}
\ee
with
\bea
G_1 &=& -\frac{J_1^2}{4} \sum_{ab} (q^{(1)}_{ab})^2-\frac{J_2^2}{4} \sum_{ab} 
(q^{(2)}_{ab})^2
-\frac{J_p^2}{2}\sum_{ab} q^{(1)}_{ab} q^{(2)}_{ab}\\
G_2 &=& - i \sum_{ab} (\hat q^{(1)}_{ab} q^{(1)}_{ab} +\hat q^{(2)}_{ab} q^{(2)}_{ab})\\
G_3 &=& \ln \left[\int \prod_a\frac{\rd u_a^{(1)} \rd  u_a^{(2)}}{2\pi/i} 
\exp\left\{- \frac{i}{2}\sum_{ab} (\lambda_\varepsilon \delta_{a,b}
-2\hat q^{(1)}_{ab})u_a^{(1)} u_b^{(1)}\right .\right . \nonumber\\
& & \left .\left . - \frac{i}{2} \sum_{ab} (\lambda_\varepsilon \delta_{a,b}- 2\hat q^{(2)}_{ab})u_a^{(2)} 
u_b^{(2)}
\right\} \right]\nonumber\\
&=& - \frac{1}{2} \ln {\rm det} (\lambda_\varepsilon\bbbone 
- 2\hat q^{(1)})- \frac{1}{2} \ln {\rm det} (\lambda_\varepsilon\bbbone - 2 \hat q^{(2)}) 
\eea
\subsubsection{Replica Symmetry --- Fixed Point Equations}

In the large $N$ limit the density of states is dominated by the saddle-point contribution
to Eq. (\ref{ZNSPInt}). Adopting the by now well-established assumptions of replica-symmetry,
and invariance under rotation in the space of replica at the relevant saddle-point
\cite{EdwJon76}
\be
q^{(\mu)}_{ab} = q_d^{(\mu)} \delta_{ab}\ ,\qquad
\hat q^{(\mu)}_{ab} = \hat q_d^{(\mu)} \delta_{ab} \ ,
\ee
one has
\bea
G&=& G_1+ G_2 + G_3 \nonumber\\
 &\simeq& n\left\{-\frac{J_1^2}{4} (q_d^{(1)})^2 - \frac{J_2^2} {4} (q_d^{(2)})^2 
- \frac{J_p^2}{2} q_d^{(1)} q_d^{(2)}  \right.\nonumber\\
 & & \left. - i \hat q_d^{(1)} q_d^{(1)} - i \hat q_d^{(2)}q_d^{(2)}
-\frac{1}{2}\left[\ln (\lambda_\varepsilon - 2\hat q_d^{(1)}) 
+\ln (\lambda_\varepsilon - 2\hat q_d^{(2)}) \right]\right\}
\label{GRS}
\eea
where terms of order $n^2$ are omitted.

Stationarity of $G$ requires that the RS order parameters solve the following fixed
point equations
\bea
-i \hat q_d^{(1)} &=& \frac{1}{2} J_1^2 q_d^{(1)} + \frac{1}{2} J_p^2 q_d^{(2)}\nonumber\\
-i \hat q_d^{(2)} &=& \frac{1}{2} J_2^2 q_d^{(2)} + \frac{1}{2} J_p^2 q_d^{(1)}\nonumber
\eea
as well as
\bea
 q_d^{(1)} &=&\frac{1}{i \lambda_\varepsilon -2 i \hat q_d^{(1)}} 
 \nonumber\\
 q_d^{(2)} &=& \frac{1}{i \lambda_\varepsilon - 2 i\hat q_d^{(2)}} \nonumber
\eea
These can be combined by eliminating the conjugate variables to give,
\be
 q_d^{(1)} = \frac{1}{i \lambda_\varepsilon + J_1^2q_d^{(1)} + J_p^2 q_d^{(2)}} \quad ,\quad 
 q_d^{(2)} = \frac{1}{i \lambda_\varepsilon + J_2^2q_d^{(2)} + J_p^2 q_d^{(1)}}
\label{CmuEq}
\ee

\subsubsection{Spectral Density for the $c\to \infty$ Case}
The average spectral density is obtained by differentiation w.r.t $\lambda$ via Eq.
(\ref{AvDOSn}). Only terms with {\em explicit\/} $\lambda$-dependence in $G$ of Eq. (\ref{GRS}) 
contribute at the saddle point. This gives
\bea
\overline{\rho(\lambda)} &=&\frac{1}{2\pi} {\rm Re}\left[q_d^{(1)}+q_d^{(2)}\right]
\eea

For the limiting cases $J_p=0$, describing two uncoupled systems, and $J_1=J_2=J$,
describing a coupling of two systems with identical statistics of intra-system couplings
we obtain the following explicit analytical results.

In the uncoupled case $J_p=0$ the solution of Eqs. (\ref{CmuEq}) is
\be
q_d^{(\mu)} = -\frac{i\lambda_\varepsilon}{2J_\mu^2} 
\pm \frac{1}{2J_\mu^2} \sqrt{4J_\mu^2-\lambda_\varepsilon^2}
\ee
giving
\be
\overline{\rho(\lambda)}=\frac{1}{2\pi}\left[ \frac{\Theta(4 J_1^2-\lambda^2)}{2J_1^2}~
\sqrt{4J_1^2-\lambda^2} + \frac{\Theta(4 J_2^2-\lambda^2)}{2 J_2^2}~
\sqrt{4 J_2^2-\lambda^2} \right]\ ,
\ee
i.e. a superposition of two independent Wigner semi-circular densities as expected.

In the case of two coupled systems which are statistically identical, with $J_1=J_2=J$,
the solution of Eqs. (\ref{CmuEq}) is $q_d^{(1)}=q_d^{(2)}=q_d$ with
\be
q_d = -\frac{i\lambda_\varepsilon}{2(J^2+J_p^2)} 
\pm \frac{1}{2(J^2+J_p^2)} \sqrt{4(J^2+J_p^2)-\lambda_\varepsilon^2}
\ee
leading to 
\be
\overline{\rho(\lambda)}=\frac{\Theta(4(J^2+J_p^2)-\lambda^2)}{2\pi(J^2+J_p^2)} ~
\sqrt{4(J^2+J_p^2)-\lambda^2} \ ,
\ee
i.e. a  Wigner semi-circular density, with the radius of the semi-circle given by
$r= 2\sqrt{J^2+J_p^2}$

In the asymmetric case $J_1\ne J_2$ we solve the fixed point equations numerically.
A non-zero coupling leads to a smoothing of the superposition of the two independent
semi-circles as anticipated; see Sec. \ref{sec:results} below.

\subsection{Two Finitely Cross-Connected Modules}
We now consider a system with the same basic setup, except that we assume the 
average number of cross-connections $c$ to be finite (also in the thermodynamic 
limit). In this case, the analysis is considerably more involved. It combines
techniques used in \cite{EdwJon76} for connected and of \cite{RodgBray88} for 
sparse random matrices, and more specifically the reformulation \cite{Ku08} 
of the sparse case that allows to proceed to explicit results. In this case the 
calculations can be carried out without restricting the distribution of the 
non-zero matrix elements of the inter-module connections to be Gaussian, and we
will in developing the theory not make that restriction.

For the average of the replicated partition function we get
\bea
\hspace{-10mm}\langle Z_N^n \rangle &\simeq& \int \prod_a\frac{\rd\vu_a^{(1)} \rd \vu_a^{(2)}}{(2\pi/i)^N} 
\exp\left\{-\frac{i}{2} \sum_a \lambda_\varepsilon (\vu^{(1)}_a \cdot \vu^{(1)}_a +
\vu^{(2)}_a \cdot \vu^{(2)}_a ) 
-\frac{J_1^2}{4N} \sum_{a,b} (\vu^{(1)}_a \cdot \vu^{(1)}_b)^2
\right .\nonumber\\
&& \left .
	   -\frac{J_2^2}{4N} \sum_{a,b} (\vu^{(2)}_a \cdot \vu^{(2)}_b)^2
 + \frac{c}{N}\sum_{ij}\left(~\left\langle\exp\left\{i K
\sum_{a} u^{(1)}_{ia}u^{(2)}_{ja}\right\}\right\rangle_K  -1 \right)
\right\}
\label{eq:avznd}
\eea
in analogy to Eq. (\ref{eq:avzn}), where $\langle ~\dots~\rangle_K$ represents 
an average over the $K_{ij}$ distribution, which is as yet left open. As in
Eq. (\ref{eq:avzn}), subdominant contributions coming from diagonal matrix 
elements are omitted.

Decoupling of sites is achieved by introducing order parameters
\be
q^{(\mu)}_{ab} =\frac{1}{N}
\sum_i u^{(\mu)}_{ia}u^{(\mu)}_{ib} ~~~~~ ,\quad \mu=1,2\ ,
\ee
as in the extensively cross-connected case before, but in addition also the replicated
densities
\be
\rho^{(\mu)}(\bm u) =\frac{1}{N}
\sum_i \prod_a \delta\Big( u_a - u^{(\mu)}_{ia}\Big) ~~~~~ ,\quad \mu=1,2\ ,
\ee
as well as the corresponding conjugate quantities. This allows to express Eq.
(\ref{eq:avznd}) as an integral over the order parameters and their conjugates,
combined with a functional integral over the replicated densities and their conjugates,
\be
\langle Z_N^n \rangle =  \int \prod_\mu \{\cD\rho^{(\mu)}\cD \hat\rho^{(\mu)}\} 
\int \prod_{\mu,a,b} \frac{\rd q^{(\mu)}_{ab}\rd \hat q^{(\mu)}_{ab}}{2\pi/N}
\exp\left\{N\left[G_1 + G_2 + G_3\right]\right\}\ ,
\label{eq:avznpathint}
\ee
with
\bea
G_1 &=& -\frac{J_1^2}{4} \sum_{ab} \big(q_{ab}^{(1)}\big)^2 
-\frac{J_2^2}{4} \sum_{ab} \big(q_{ab}^{(2)}\big)^2 \nonumber\\
& & + c\int \rd \rho^{(1)}(\bm u) 
\rd \rho^{(2)}(\bm v) \left(\left\langle\exp\left\{i K
\sum_{a} u_{a} v_{a}\right\}\right\rangle_K  -1\right)\nonumber\\
G_2 &=&  -i\sum_{ab}\Big(\hat q_{ab}^{(1)}q_{ab}^{(1)} + \hat q_{ab}^{(2)}q_{ab}^{(2)}\Big)
- i \int \rd \bm u\Big(\hat\rho^{(1)}(\bm u)\rho^{(1)}(\bm u) 
+ \hat\rho^{(2)}(\bm u)\rho^{(2)}(\bm u) \Big)\nonumber\\
G_3 &=& \ln \int \prod_a \rd u_a \exp\left\{i\hat\rho^{(1)}(\bm u) - \frac{i}{2} 
\sum_{ab}\Big(\lambda_\varepsilon \delta_{ab} - 2\hat q_{ab}^{(1)}\Big) u_a u_b \right\}
\nonumber\\
   & & + \ln \int \prod_a \rd u_a \exp\left\{i\hat\rho^{(2)}(\bm u) - \frac{i}{2}
\sum_{ab}\Big(\lambda_\varepsilon \delta_{ab} - 2\hat q_{ab}^{(2)}\Big) u_a u_b \right\}
\nonumber
\eea
Here we have introduced abbreviations of the form $\rd \rho^{(\mu)}(\bm u) \equiv
\rd \bm u \,\rho^{(\mu)}(\bm u)$ for integrals over densities where appropriate to 
simplify notation.
\subsubsection{Replica Symmetry \& Self-Consistency Equations}
Eq. (\ref{eq:avznpathint}) is evaluated by the saddle point method. The saddle point
for this problem is expected to be replica-symmetric and symmetric under rotation in
the space of replica as in the extensively cross-connected case. In the present context
this translates to an ansatz of the form 
\be
q^{(\mu)}_{ab} = q_d^{(\mu)} \delta_{ab}\ ,\qquad
\hat q^{(\mu)}_{ab} = \hat q_d^{(\mu)} \delta_{ab}\ ,
\ee
for the Edwards Anderson type order parameters \cite{EdwJon76} and
\bea
\rho^{(\mu)}(\bm u) &=& \int \rd \pi^{(\mu)}(\omega) \prod_a 
\frac{\exp\big[-\frac{\omega}{2} u_a^2\big]}{Z(\omega)}\ ,
 \nonumber\\
i\hat\rho^{(\mu)}(\bm u) &=& \hat c^{(\mu)} \int \rd \hat\pi^{(\mu)}(\hat{\omega}) 
\prod_a \frac{\exp\big[-\frac{\hat\omega}{2}u_a^2 \big]}{Z(\hat{\omega})}\ ,
\eea
i.e. an uncountably infinite superposition of complex Gaussians (with $\mathrm{Re}
\omega \ge 0$ and $\mathrm{Re}\hat\omega \ge 0$) for the replicated densities and 
their conjugates \cite{Ku08}. Here we have introduced the shorthand
\be
Z(\omega) = \int \rd u \exp\left[-\frac{\omega}{2} u^2\right]
= \sqrt{2 \pi/\omega}\ .
\label{defZ}\\
\ee
The $\hat c^{(\mu)}$ in the expressions for $\hat \rho^{(\mu)}$ are to be determined 
such  that the densities $\hat\pi^{(\mu)}$ are normalised. 

This ansatz translates path-integrals over the replicated densities $\rho$ and 
$\hat\rho$ into path-integrals over the densities $\pi$ and $\hat\pi$, giving
\be
\langle Z_N^n \rangle= \int \prod_\mu \{\cD\pi^{(\mu)}\cD \hat\pi^{(\mu)}\} 
\int \prod_{\mu} \frac{\rd q^{(\mu)}_{d}\rd \hat q^{(\mu)}_{d}}{2\pi/N}
\exp\left\{N \left[G_1 + G_2 + G_3\right]\right\}
\ee
with now
\bea
G_1 &\simeq & n\left[-\frac{J_1^2}{4}(q_d^{(1)})^2 -\frac{J_2^2}{4} (q_d^{(2)})^2  
\right .
\nonumber\\
&&\left . + c \int\rd \pi^{(1)}(\omega)\rd\pi^{(2)}(\omega') \left\langle 
\ln \frac{Z_2( \omega,\omega',K)}{Z(\omega)Z(\omega')}\right\rangle_K \right]\ , \\
G_2 &\simeq & - n\left[i\hat q_d^{(1)} q_d^{(1)}  +i \hat q_d^{(2)} q_d^{(2)}\right]
\nonumber\\
& & - \sum_{\mu=1}^2\left[\hat c^{(\mu)}+ n \hat c^{(\mu)}\int \rd \hat\pi^{(\mu)}
(\hat{\omega}) \rd\pi^{(\mu)}(\omega) \ln \frac{Z(\hat{\omega}+\omega)}
{Z(\hat{\omega})Z(\omega)}\right]\ ,\\
G_3 &\simeq & \sum_{\mu=1}^2\left[\hat c^{(\mu)} + n
\sum_{k=0}^\infty p_{\hat c^{(\mu)}}(k) \int \{\rd \hat\pi^{(\mu)}\}_k 
\ln \frac{Z^{(\mu)}(\{{\hat\omega}\}_k)}{\prod_{\ell=1}^k Z({\hat\omega_\ell})}
\right]\ .
\eea
Here $\{\rd \hat\pi^{(\mu)}\}_k \equiv \prod_{\ell=1}^k \rd \hat\pi^{(\mu)}(\hat 
\omega_\ell)$, while $\{\hat\omega\}_k\equiv \sum_{\ell=1}^k \hat\omega_\ell$, 
and 
\be
p_{\hat c^{(\mu)}}(k) = \frac{\hat c^{(\mu)k}}{k!} \exp[-\hat c^{(\mu)}]
\ee
is a Poissonian distribution with average $\langle k\rangle =\hat c^{(\mu)}$, and we 
have introduced the partition functions
\bea
Z^{(\mu)}(\{\hat\omega\}_k)  &=& \int \frac{\rd u}{\sqrt{2\pi/i}}~
\exp\left[-\frac{1}{2}\bigg(i\lambda_\varepsilon - 2 i\hat q_d^{(\mu)} + 
\{\hat\omega\}_k \bigg) u^2 \right] \nonumber\\
&=& \left(\frac{i}{i \lambda_\varepsilon- 2 i\hat q_d^{(\mu)}+ \{\hat\omega\}_k}\right)^{1/2}
\label{defZmu}\ ,\\
 Z_2(\omega,\omega',K) &=& \int \rd u \rd v ~
\exp\left[-\frac{1}{2}\bigg(\omega u^2 + \omega' v^2 - 2i K u v\bigg)\right] 
= \frac{2\pi}{\sqrt{\omega \omega' + K^2}}\ .
\label{defZ2}
\eea

The stationarity conditions for $\pi^{(1)}(\omega)$ and $\pi^{(2)}(\omega)$ then
read 
\be
 \hat c^{(1)}\int \rd \hat\pi^{(1)}(\hat{\omega})
\ln \frac{Z(\hat{\omega}+\omega)} {Z(\hat{\omega})Z(\omega)}
=c \int\rd\pi^{(2)}(\omega')
\left\langle \ln \frac{Z_2(\omega,\omega',K)}{Z(\omega)Z(\omega')}\right\rangle_K
+ \mu_1
\label{statpi1}
\ee
and
\be
\hat c^{(2)}\int \rd \hat\pi^{(2)}(\hat{\omega})
\ln \frac{Z(\hat{\omega}+\omega)} {Z(\hat{\omega})Z(\omega)}
=c \int\rd\pi^{(1)}(\omega')
\left\langle\ln \frac{Z_2(\omega',\omega,K)}{Z(\omega)Z(\omega')}\right\rangle_K
+ \mu_2\ , 
\label{statpi2}
\ee
with $\mu_1$ and $\mu_2$ Lagrange multipliers to take the normalisation of $\pi^{(1)}$ 
and $\pi^{(2)}$ into account. The stationarity conditions for the $\hat\pi^{(\mu)}
(\hat{\omega})$ are
\be
\hspace{-20mm}
c\int \rd\pi^{(\mu)}({\bm\omega}) \ln \frac{Z(\hat{\omega}+\omega)} 
{Z(\hat{\omega})Z(\omega)} =\sum_{k\ge 1} k~p_{\hat c^{(\mu)}}(k)
\int \{\rd \hat\pi^{(\mu)}\}_{k-1} 
\ln \frac{Z^{(\mu)}(\hat{\omega}+\{\hat\omega\}_{k-1})}
{Z(\hat{\omega})\prod_{\ell=1}^{k-1} Z(\hat\omega_\ell)} +  \sigma_\mu
\label{stathatpi}
\ee
where the $\sigma_\mu$ are once more Lagrange multipliers to take the normalisation of
the $\hat\pi^{(\mu)}(\hat{\omega})$ into account.

The stationarity conditions for the $q_d^{(\mu)}$ are
\be
  i\hat q_d^{(\mu)} = -\frac{J_\mu^2}{2} q_d^{(\mu)} \ ,
\label{statqqd}
\ee
the corresponding ones for the conjugate variables give 
\bea
q_d^{(\mu)} &=&  \sum_{k=0}^\infty p_{\hat c^{(\mu)}}(k) \int \{\rd
\hat\pi^{(\mu)}\}_k  ~\left\langle u^2 \right\rangle^{(\mu)}_{\{{\hat\omega}\}_k} 
\nonumber  \\
& = & \sum_{k=0}^\infty p_{\hat c^{(\mu)}}(k) \int \{\rd
\hat\pi^{(\mu)}\}_k  ~\frac{1}{i \lambda_\varepsilon + J_\mu^2 q_d^{(\mu)}+ \{\hat\omega\}_k}
\label{stathatqd}
\eea
in which $\left\langle \dots \right\rangle^{(\mu)}_{\{{\hat\omega}\}_k}$ is defined as 
an average w.r.t. the Gaussian weight in terms of which $Z^{(\mu)}(\{\hat\omega\}_k)$
is defined. We have used Eq. (\ref{statqqd}) to express the $i\hat q_d^{(\mu)}$ in terms 
of the $q_d^{(\mu)}$.

Following \cite{Mon98,MezPar01}, the stationarity conditions for $\pi^{(1)}(\omega)$ 
and $\pi^{(2)}(\omega)$ are  rewritten in a form that suggests solving 
them via a population based algorithm.  In the present case we get \cite{Ku+07,Ku08}
\be
\hat\pi^{(1)}(\hat{\omega})= \int\rd\pi^{(2)}(\omega') 
\left\langle\delta\Big(\hat{\omega} - \hat{\Omega}(\omega',K)\Big)\right\rangle_K
\label{eq:hatpi1}
\ee
and
\be
\hat\pi^{(2)}(\hat{\omega})= \int\rd\pi^{(1)}(\omega') 
\left\langle\delta\Big(\hat{\omega} - \hat{\Omega}(\omega',K)\Big)\right\rangle_K\ ,
\label{eq:hatpi2}
\ee
in which 
\be
\hat\Omega = \frac{K^2}{\omega'}\ ,
\ee
while
\be
\pi^{(\mu)}({\omega})= \sum_{k\ge 1} \frac{k}{c} p_c(k)
\int \{\rd \hat\pi^{(\mu)}\}_{k-1} \delta\left(\omega - \Omega^{(\mu)}_{k-1}\right)
\label{eq:pimu}
\ee
with
\be
\Omega^{(\mu)}_{k-1} = i\lambda_\varepsilon + J_\mu^2 q_d^{(\mu)} + \sum_{\ell=1}^{k-1} 
\hat{\omega}_{\ell}\ .
\ee

In Eqs. (\ref{eq:hatpi1}), (\ref{eq:hatpi2}), and (\ref{eq:pimu}), we have already invested
$\hat c^{(\mu)} = c$ to enforce that the $\hat\pi^{(\mu)}$ are normalized.

For the spectral density we obtain the same formal result as for the extensively 
cross-connected system before,
\be
\overline{\rho(\lambda)}= \frac{1}{2\pi}{\rm Re}\left[q_d^{(1)}+q_d^{(2)}\right]\ .
\ee

The solution of these coupled sets of equations is considerably more involved than for
the extensively cross-connected systems considered earlier, as it involves solving 
equations for the coupled macroscopic order parameters $q_d^{(1)}$ and $q_d^{(2)}$, 
which are themselves expressed in terms of averages over self-consistent solutions of a 
pair of non-linear integral equations parameterised by these order parameters. However, we
have found that a population dynamics in which values of $q_d^{(1)}$ and $q_d^{(2)}$ are
iteratively updated using Eq. (\ref{stathatqd}) by sampling from the corresponding
populations rapidly converges to the correct solution.

\section{The Multi-Modular Case}
Finally, we consider systems with $M$ modules, each of size $N$, mutually cross-connected
with  finite connectivity. Inside modules we may or may not have all-to-all Gaussian 
couplings of variances $J_\mu^2/N$. In addition there are finitely many $\cO(1)$ 
module-to-module couplings  for each vertex, with average connectivities $c^{\mu\nu}$
for couplings between nodes in modules $\mu$ and $\nu$ (possibly including the case
$\mu=\nu$). Evaluating this case requires a fairly straightforward generalisation of the
setup developed earlier. Here we only produce the fixed point equations and the final
expression for the average spectral density.
We get
\be
\hat\pi^{(\mu)}(\hat\omega)= \sum_{\nu}
\frac{c^{\mu\nu}}{c^{\mu}} \int\rd\pi^{(\nu)}(\omega) 
\left\langle\delta\Big(\hat\omega - \hat\Omega(\omega,K)\Big)\right\rangle_{\mu\nu} 
\label{eq:cphatpi}
\ee
and
\be
\pi^{(\mu)}(\omega)= \sum_{k\ge 1} \frac{k}{c^\mu} p_{c^\mu}(k)
\int \{\rd \hat\pi^{(\mu)}\}_{k-1} \delta\left(\omega -
\Omega^{(\mu)}_{k-1}\right)\ ,
\label{eq:cppi}
\ee
as well as
\be
q_d^{(\mu)} = \sum_{k=0}^\infty p_{c^\mu}(k) \int \{\rd
\hat\pi^{(\mu)}\}_k ~\frac{1}{\Big(i\lambda_\varepsilon + J_\mu^2 q_d^{(\mu)} +
\sum_{\ell=1}^{k} \hat{\omega}_{\ell}\Big)}
\label{eq:cpqd}
\ee
with
\be
\hat\Omega (\omega,K) = \frac{K^2}{\omega}\ ,
\label{eq:cphatom}
\ee
and
\be
\Omega^{(\mu)}_{k-1} = i\lambda_\varepsilon + J_\mu^2 q_d^{(\mu)} +\sum_{\ell=1}^{k-1}
\hat{\omega}_{\ell}\ .
\label{eq:cpom}
\ee

In Eq. (\ref{eq:cphatpi}), $\langle \dots \rangle_{\mu\nu}$ denotes an average over 
the distribution
of couplings connecting modules $\mu$ and $\nu$, and we have the
normalisation  $\sum_\nu
  c^{\mu\nu}= c^{\mu}$.

For the spectral density we obtain the (obvious) generalisation of those obtained for 
the two modules case before, viz.
\be
\overline{\rho(\lambda)}= \frac{1}{M\pi}{\rm Re}\sum_{\mu=1}^M q_d^{(\mu)}\ .
\label{eq:cprho}
\ee
Generalising these to the situation of varying module-sizes $N_\mu = r_\mu N$, with 
$r_\mu >0$, is straightforward.

\section{Results}
\label{sec:results}

\subsection{Extensively Cross-Connected Systems}

The results we obtain for extensively cross-connected systems can be understood in 
terms of superpositions of Wigner semi-circles one would expect for the spectral
densities of the modules if they were uncoupled, but smoothed (and broadened) by 
the interaction. We have checked that our results agree perfectly with simulations,
but have not included results of simulations in the figures for the extensively
cross-connected systems below.

\begin{figure}
\epsfig{file=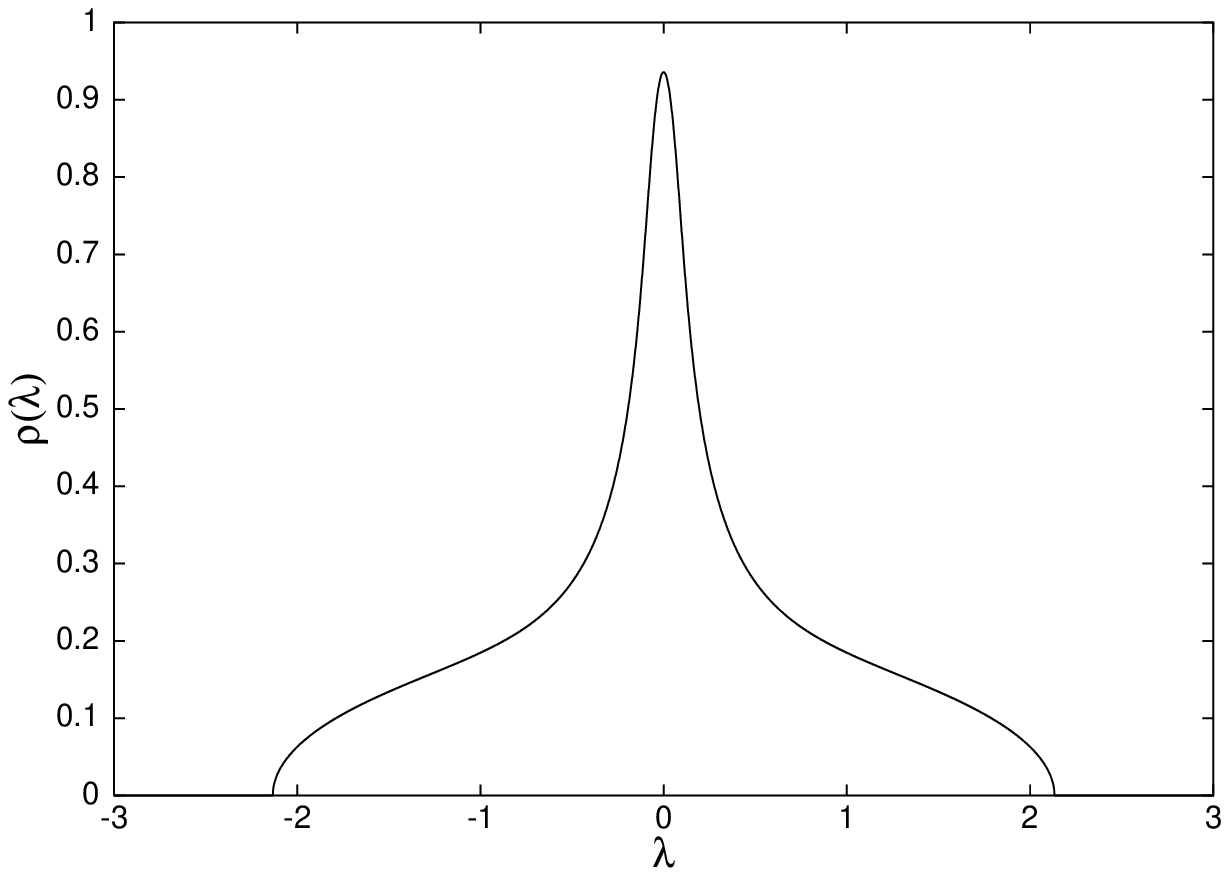,width=0.495\textwidth}\hfill
\epsfig{file=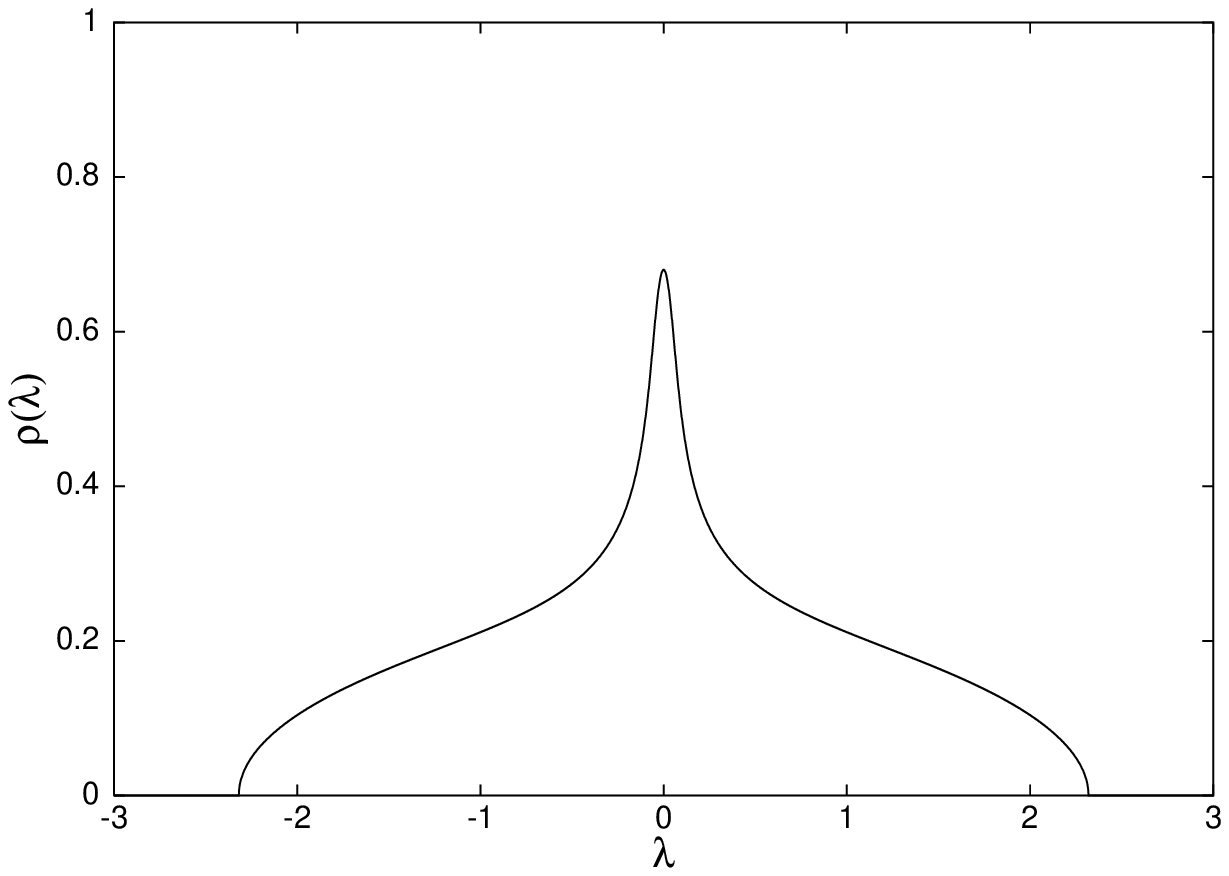,width=0.495\textwidth}
\caption{Spectral densities for $J_1=1$, $J_2=0.1$, and $J_p=0.5$ (left), and $J_p=0.75$
(right).}
\end{figure}

In Fig. 1 we explore the effect of the inter-module coupling strength $J_p$ on the
spectral density of a system of two coupled modules with intra-module coupling 
strengths $J_\mu$ differing by one order of magnitude. The reader is invited to 
compare the results with those expected for in a situation where the two modules
were non-interacting, namely a simple superposition of semi-circular densities of
radii $2J_1$ and $2J_2$, respectively.

In Fig. 2 we keep the strength of the inter-module couplings but vary the ratio of 
the two intra-module couplings.

Results for uncoupled modules or for coupled modules with identical intra-module 
coupling statistics are not shown; they were found to be in perfect agreement 
with the simple analytical results presented in Sec. 2.1.2.

\begin{figure}
\epsfig{file=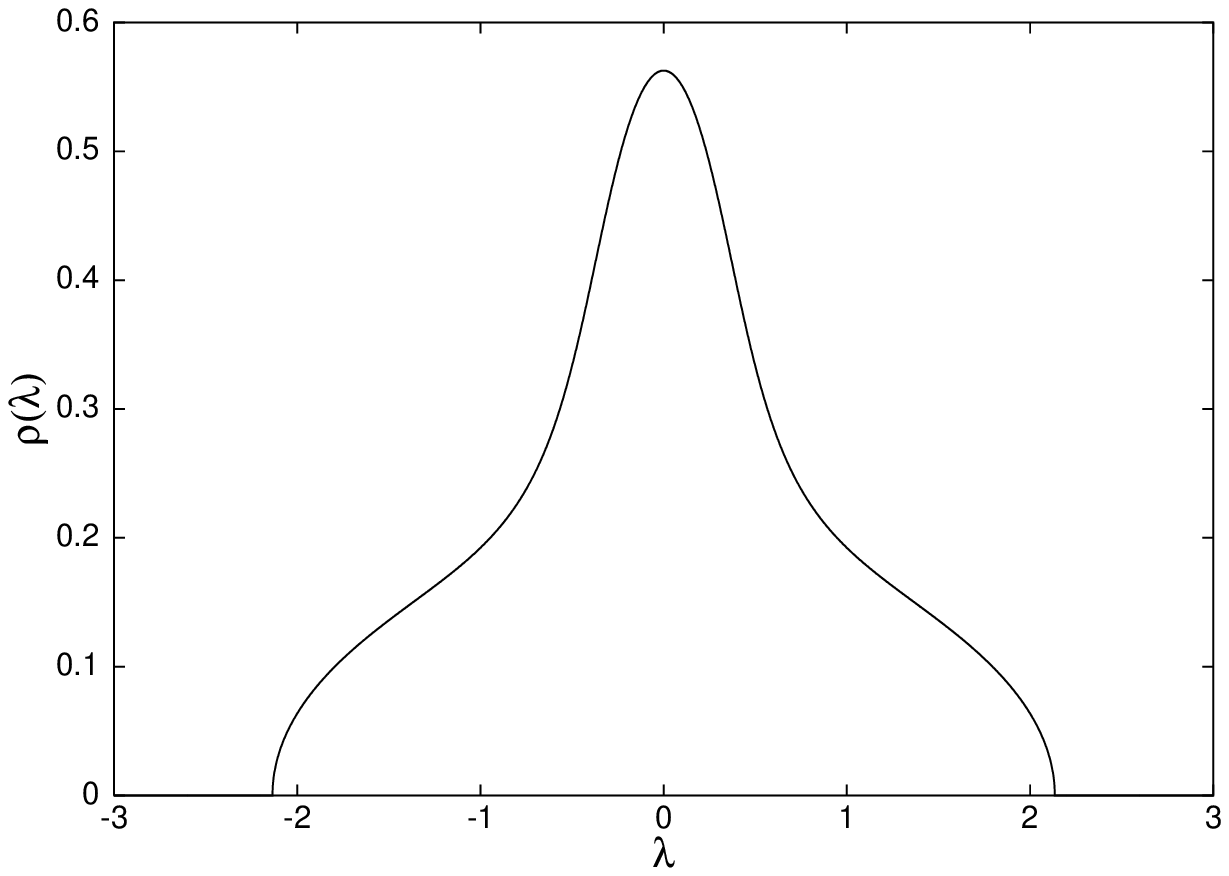,width=0.495\textwidth}\hfill
\epsfig{file=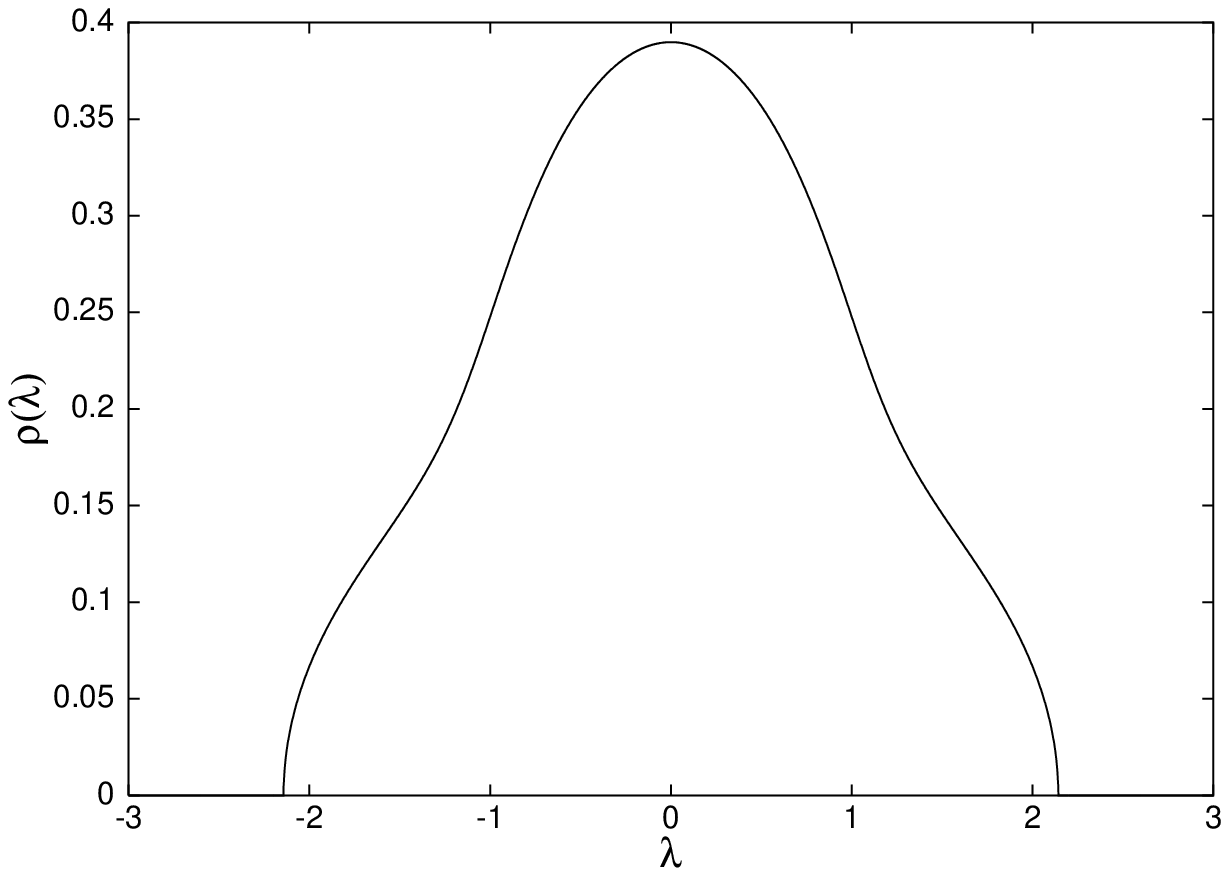,width=0.495\textwidth}
\caption{Spectral densities for $J_1=1$, $J_2=0.25$ (left), and $J_2=0.5$ (right),
and $J_p=0.5$ in both cases}
\end{figure}

\subsection{Finitely Cross-Connected Systems}

The modifications generated by cross-connections that remain finite in number and 
strength for each node in each of the blocks are more pronounced than those created
by extensive (infinitesimal) cross-connections. We mention just two fairly drastic
modifications. First, the spectral density of a system of finitely cross-connected
modules does no longer have sharp edges with square-root singularities of the spectral
density at the edges as in the case of extensively cross-connected systems, where it
derives from a perturbed superposition of semi-circular densities, but rather tails 
with decay laws that depend on the nature of the distribution of cross-connections. 
In the present case of Poisson distributions of cross-connections we find these 
tails to exhibit exponential decay. Second, even with identical intra-module coupling 
statistics the spectral density is no longer given by a simple semi-circular law as 
it is for the extensively cross-connected case, but rather exhibits marked deviations 
from the semi-circular law, with details depending both on the number and the strengths
of the cross-connections, as shown in Fig. 3.

As the numerics in the present finitely cross-connected case is considerably more
involved we present results of the analytic theory together with checks against 
numerical diagonalization. Figs 3-4 demonstrate that the analytic results are in
excellent agreement with numerical simulations, virtually indistinguishable for the
parameters and the statistics used.

\begin{figure}
\epsfig{file=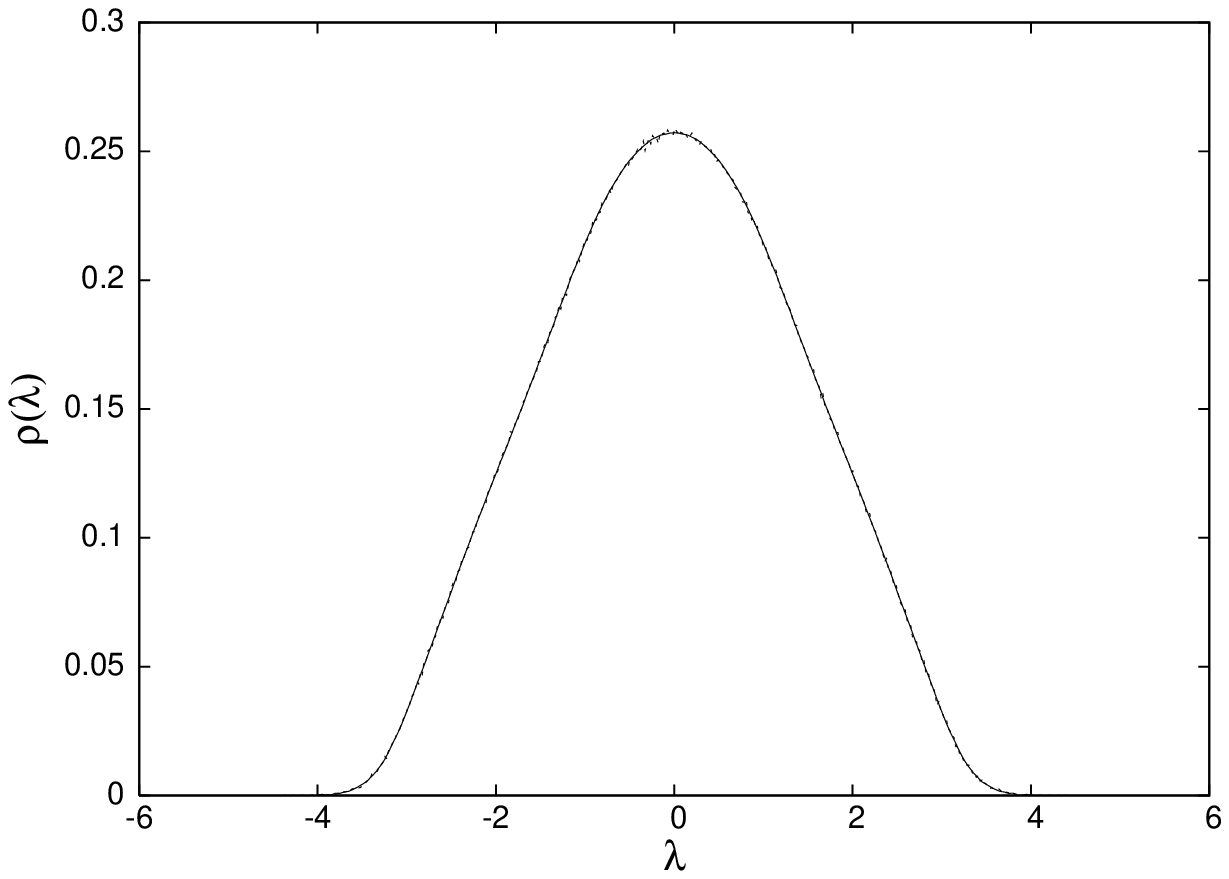,width=0.495\textwidth}\hfill
\epsfig{file=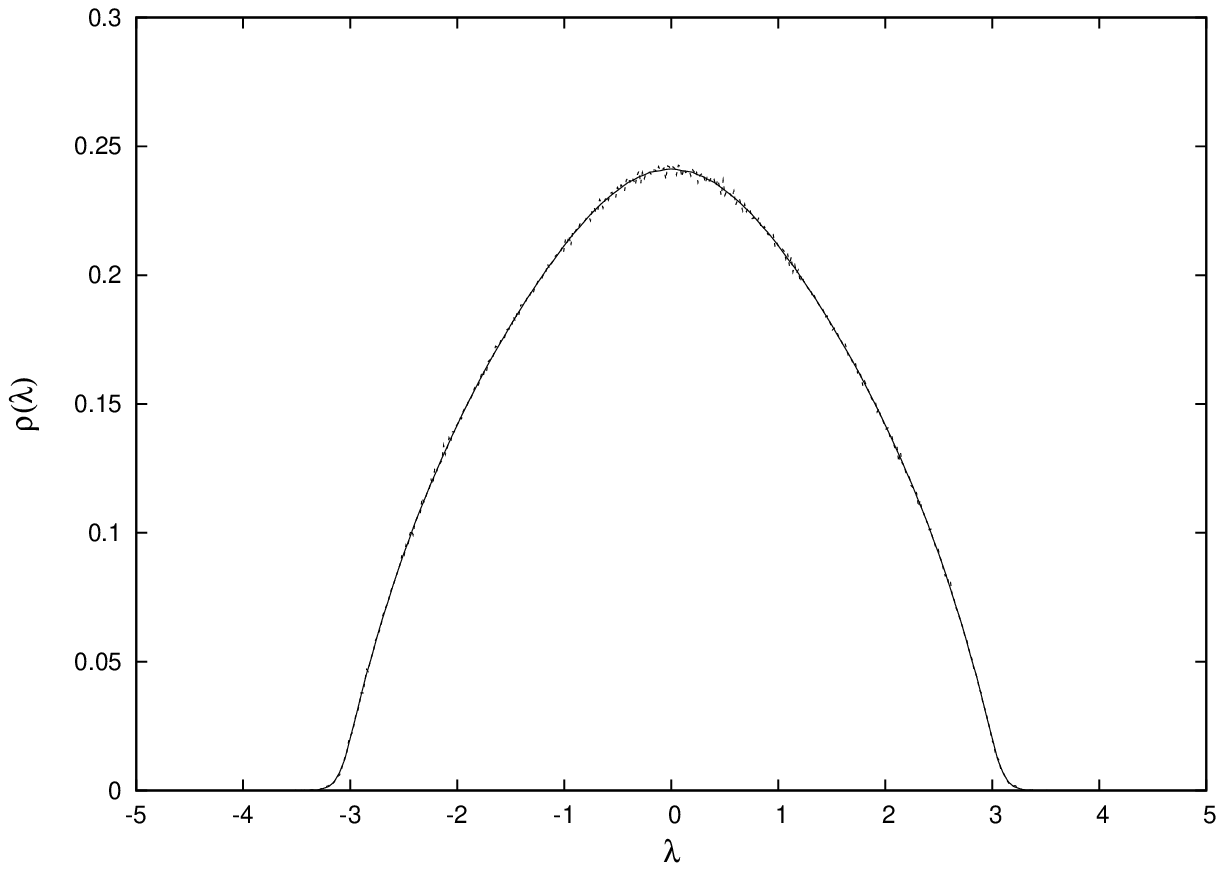,width=0.495\textwidth}
\caption{Spectral densities for $J_1=J_2=1 $ and $J_p= 1/\sqrt c$ 
for $c=2$ (left) and $c=5$ (right). Results of numerical diagonalizations of 500
matrices containing two coupled blocks, each of dimension 1000, are shown for 
comparison (dashed lines); they are virtually indistinguishable from results of
the analytic theory.}
\end{figure}

Fig 3 exhibits spectral densities of a two-module system with identical  intra-module
coupling statistics, and Gaussian cross-connections with a Poissonian degree statistics,
of average cross-coordinations 2 and 5 respectively. In the first case the spectral
density has a shape close to triangular, though more rounded at the tip than what is
known from spectral density of certain scale-free systems.

\begin{figure}
\begin{center}
\epsfig{file=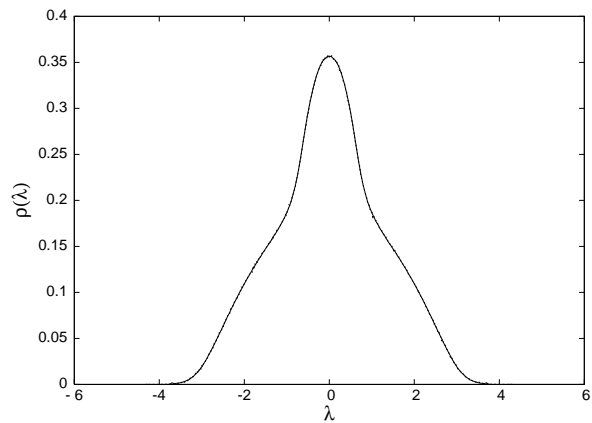,width=0.495\textwidth}
\end{center}
\caption{Spectral density for $J_1=1$, $J_2=0.5$ and $J_p= 1/\sqrt c$ 
for $c=2$ (full line). Results of numerical diagonalizations of 500 matrices containing 
two coupled blocks, each of dimension 1000, are shown for comparison (dashed line), and
are once more virtually indistinguishable from results of the analytic theory.}
\end{figure}

Fig. 4 looks at a system of two modules having intra-module connections of different 
strengths, and average cross-coordinations 2; results resemble the corresponding case 
of extensive cross-connectivity, apart from the exponential tails, which are an 
exclusive feature of the finitely cross-connected case.

\section{Summary and Conclusions}

We have evaluated spectral densities of symmetric matrices describing modular 
systems. Modularity is regarded as one of several routes to create heterogeneity
in interacting systems. In some biological systems in fact, modularity of interactions
appears to be a natural consequence of compartmentalization; systems with cellular
structure, or sub-structures within cells  come to mind, where heterogeneity of
interaction patterns due to modularity of the system would seem to enjoy a greater
degree of plausibility than heterogeneity as observed in certain scale free systems.

In any case, whenever large systems with different levels of organization are considered,
modularity appears to be a feature to be reckoned with.

While the results for extensively cross-connected systems can be understood in
terms of perturbed superpositions of semi-circular spectral densities characteristic
of the individual modules if in isolation, the case of finitely cross-connected is
more sensitive to details of the statistics of cross-connections. 

We have evaluated examples only for systems containing two coupled modules, but 
presented the general theory for multi-modular systems in Sec 3. Our derivations and 
results were restricted to the case where entries of the connectivity matrix were 
sampled independently (leading to Poissonian degree distributions of inter-module
connections in the finitely cross-connected case). Unlike in the case of a single
module \cite{Ku+07}, the case where degree distributions of inter-module connections 
differ from Poissonian distributions requires more substantial modifications to the 
structure of the theory We will deal with these, as well as with spectra of matrices 
corresponding to small-world systems \cite{WaStro98} in a separate publication 
\cite{KuVM09}.

\section*{References}
\bibliography{../../MyBib}
\end{document}